\documentclass[prc,twocolumn,preprintnumbers,superscriptaddress,showpacs,nofootinbib]{revtex4-1}
\usepackage{graphicx,times}
\usepackage{amsmath,amsfonts,amssymb}

\begin{document}

\title{Looking for bound states and resonances in the $\eta^\prime K\bar K$ system}
\author{A. Mart\'inez Torres}

\affiliation{Instituto de F\'isica, Universidade de S\~ao Paulo, 05389-970 S\~ao 
Paulo, SP, Brazil.}
\author{K.~P.~Khemchandani}
\affiliation{ Faculdade de Tecnologia, Universidade Estadual de Rio de Janeiro, 27537-000, Resende, Rio de Janeiro}

\date{\today}

\begin{abstract}
Motivated by the continuous experimental investigations of $X(1835)$ in three-body decay channels like $\eta^\prime \pi^+ \pi^-$, we investigate the $\eta^\prime K \bar K$ system with the aim of searching for bound states and/or resonances when the dynamics involved in the $K\bar K$ subsystem can form the resonances: $f_0(980)$ in isospin 0 or $a_0(980)$ in isospin 1. For this, we solve the Faddeev equations for the three-body system. The input two-body $t$-matrices are obtained by solving Bethe-Salpeter equations in a coupled channel formalism. As a result, no signal of a three-body bound state or resonance is found.  
\end{abstract}

\maketitle
An observation of a resonance-like structure around 1830 MeV, $X(1835)$, has been reported in several processes, with the recent-most finding being in the mass spectrum of $\eta^\prime \pi^+ \pi^-$ by the BES collaboration~\cite{Ablikim:2010au}. The first observation of $X(1835)$ in the  $\eta^\prime \pi^+ \pi^-$ mass spectrum, in the process $J/\psi \to \gamma \eta^\prime \pi^+ \pi^-$, was discussed in Ref.~\cite{Ablikim:2005um}, where a Breit-Wigner fit to the data yielded a mass $M = 1833.7 \pm 6.1 \pm 2.7$ MeV a width $\Gamma = 67.7 \pm 20.3 \pm 7.7$ MeV.   The same process is studied with a larger statistics by BESIII in Ref.~\cite{Ablikim:2010au} where, apart from the confirmation of $X(1835)$, the finding of two new states is reported: $X(2120)$ and $X(2370)$. A more recent analysis of the $\eta^\prime \pi^+ \pi^-$ data \cite{Ablikim:2016itz}, focussed on the energy region of $X(1835)$, shows that a fit to the data in this region requires either the presence of a much broader state ($\Gamma \sim 247$ MeV), distorted by the cusp of $p \bar p$, or an interference between a  broad and a narrow state. The fit shows that the broad state, in any case, couples strongly to the $p \bar p$ system~\cite{Ablikim:2016itz}. An enhancement near the $p \bar p$ threshold in the BES data has been found in some processes (like $J/\psi \to \gamma p \bar p$, $\psi(2s) \to \gamma p \bar p$ \cite{BESIII:2011aa}) but not in some other processes (like, $J/\psi \to \omega p \bar p$ \cite{Ablikim:2013cif}, $J/\psi \to \phi  p \bar p$ \cite{Ablikim:2015pkc}). The decay of $\psi(2s)$ has been studied by the CLEO Collaboration also, but the data shows no $p \bar p$ threshold enhancements in the mass spectra of $\gamma p \bar p$, $\pi^0 p \bar p$ and $ \eta p \bar p$ \cite{Alexander:2010vd}. All these findings have generated a series of discussions on the possibility of the existence of a baryonium, or other alternative explanations of the enhancement seen around 1830 MeV \cite{Dmitriev:2013xla,Deng:2013aca,Kang:2015yka,Samart:2012gk,Deng:2012wi,Liu:2009vm,Wang:2006sna,Abud:2009rk,Dedonder:2009bk,Chen:2010an,Rosner:2006vc,Entem:2006dt}. A resonance like structure around 1830 MeV is also found in the mass spectrum of  $\eta K \bar K$ \cite{Ablikim:2015toc}, $\eta \pi^+ \pi^-$, where the $K \bar K$ is found to come dominantly from $f_0(980)$ in the former case.  It is not clear if all the states found around 1830 MeV in different systems are the same and the origin of this/these state(s) is still an open question. In the present manuscript we study the possibility of understanding $X(1835)$ as a bound state arising from three pseudoscalar dynamics involving the $\eta^\prime$-meson.

The dynamics of a system of pseudoscalar mesons is related to the low energy regime of the Quantum Chromodynamics (QCD), which can be described in terms of the chiral perturbation theory ($\chi\text{PT}$).
The latter is an effective field theory based on the fact that the QCD Lagrangian with massless $u$, $d$ and $s$ quarks has an SU(3)$_R$$\times$ SU(3)$_{L}$
chiral symmetry. This symmetry is spontaneously broken to SU(3)$_V$, giving rise to an octet of Goldstone bosons, which are identified
with the octet formed by the pseudoscalar mesons: $\pi$, $K$ and $\eta$. Particles which become massless in the chiral limit of zero quark masses, $m_{u,\,d,\,s}\to 0$. 
The ninth pseudoscalar, the $\eta^\prime$ meson, which was found independently, but almost
at the same time, by two collaborations~\cite{kalb,gold} in 1964, is an interesting hadron: it is closely related to the axial $U_A(1)$ anomaly~\cite{wein,thooft,witten}.
This fact prevents the $\eta^\prime$-meson to become massless even in the chiral limit. Thus, the $\eta^\prime$-meson is not included explicitly in the
Lagrangian in the conventional $\chi\text{PT}$.

A way to incorporate $\eta^\prime$, however, could be inspired by the works of Witten, `t Hooft and others~\cite{thooft,witten}, who showed that in the limit of infinite number of colors ($N_c\to\infty$) of QCD the SU(3) singlet state, $\eta_1$, is massless and the global SU(3)$_R$ $\times$ SU(3)$_{L}$ symmetry is replaced by U(3)$_R$$\times$ U(3)$_{L}$. This is because in the large $N_c$ limit the anomaly related to the axial current is $1/N_c$ suppressed. This fact can be used to incorporate $\eta^\prime$ in an effective field theory based on chiral symmetry, since $\eta_1$ becomes the ninth Goldstone boson and can be included in an extended U(3)$_R$$\times$ U(3)$_{L}$ chiral Lagrangian (see, for example,~\cite{herrera,kaiser,jamin} for more details). Alternative approaches to include the singlet state in an effective field theory have also been developed~\cite{bora1,bora2}. 

Thus, to build a Lagrangian based on chiral symmetry and including at the same time the $\eta^\prime$ meson, in the spirit of  Refs.~\cite{thooft,witten,herrera,kaiser,jamin,bora1,bora2}, the physical $\eta$ and $\eta^\prime$ fields are introduced as the admixtures of the SU(3) singlet $\eta_1$ and octet $\eta_8$ states. Indeed, the $\eta-\eta^\prime$ mixing has received a lot of attention in the recent past. Usually, within the mixing scheme, the $\eta$ and $\eta^\prime$ mesons are considered as linear combinations of $\eta_1$ and $\eta_8$ through a mixing angle $\theta$
\begin{align}
|\eta\rangle&=\text{cos}\theta~|\eta_8\rangle-\text{sen}\theta~|\eta_1\rangle,\nonumber\\
|\eta^\prime\rangle&=\text{sen}\theta~|\eta_8\rangle+\text{cos}\theta~|\eta_1\rangle.\label{mix}
\end{align}
 The values obtained for this mixing angle range, typically, from $-13^\circ$ to $-22^\circ$. These values are extracted, just to mention a few examples, from the decays of $\eta$ and $\eta^\prime$ to two photons, decays of $J/\psi$, etc.~\cite{gilman,akhoury,bramon1,venugopal}. Considering this mixing angle, the SU(3) matrix containing the Goldstone bosons can be extended to U(3) as
 \begin{widetext}
 \begin{align}
 \phi&=\left(\begin{array}{ccc} \frac{1}{\sqrt{2}}\pi^0+\frac{1}{\sqrt{3}}\eta+\frac{1}{\sqrt{6}}\eta^\prime&\pi^+&K^+\\
 \pi^-&-\frac{1}{\sqrt{2}}\pi^0+\frac{1}{\sqrt{3}}\eta+\frac{1}{\sqrt{6}}\eta^\prime&K^0\\
 K^-&\bar{K}^0&-\frac{1}{\sqrt{3}}\eta+\frac{2}{\sqrt{6}}\eta^\prime
 \end{array}\right),\label{phi}
 \end{align}
 \end{widetext}
 where the standard $\eta-\eta^\prime$ mixing is considered [Eq.~(\ref{mix}) with $\text{sin}\theta=-1/3$, thus $\theta\sim -20^\circ$]. Also, a two-mixing angle scheme has been proposed~\cite{kaiser1,kaiser2} and adopted to explain some decay widths of the $\eta$ and $\eta^\prime$ mesons, radiative decays, pseudoscalar decay constants, and other quantities~\cite{feldmann,bramon2}. We stick to the approach with one mixing angle.

Using the matrix in Eq.~(\ref{phi}), at leading order in large $N_c$, the lowest order Lagrangian describing the interaction between two pseudoscalar mesons is given by~\cite{herrera,kaiser,jamin,guo}
\begin{align}
\mathcal{L}=\frac{1}{12 f^2}\langle (\partial_\mu\phi\phi-\phi\partial_\mu\phi)^2+M\phi^4\rangle,\label{L}
\end{align}
with $M=\text{diag}(m^2_\pi,m^2_\pi,2m^2_K-m^2_\pi)$.

The interaction of the $\eta^\prime$ meson with other pseudoscalars in S-wave is rather weak, and neither bound state nor resonance have been found theoretically due to this dynamics. However, it was shown in Refs.~\cite{jamin,guo} that inclusion of $\eta^\prime$ in the coupled channel analysis is required  to  reproduce the isospin $I=1/2$ and $I=3/2$ S-wave $K\pi$ phase-shift up to energies of 1.3 GeV. 
In fact, a pole around 700 MeV with a width near 600 MeV is found and identified with the $\kappa$ resonance in Refs~\cite{jamin,guo}. Note, however, that the presence of the $\eta^\prime K$ channel, although being important for the reproduction of the data around 1.3 GeV, is not essential for the understanding of the properties and nature of the $\kappa$ resonance~\cite{oset,pelaez,eef}.

Contrary to the weakness of the $\eta^\prime$ interaction with other pseudoscalars, the S-wave  interaction of systems like $K \bar{K}$ is known to be strong, and generates poles related to the $f_0(980)$ and $a_0(980)$ resonances~\cite{eef,oset,pelaez}.  It is then plausible that in a system like $\eta^\prime K \bar K$  the strong attraction in the $K \bar K$ system could be enough, together with a weak interaction in the subsystems having a $\eta^\prime$, to generate a state with a three-body nature. Such a plausibility should not be surprising because the three-body dynamics is more complex and richer than the one associated with a two-body system, and states of three-body nature can be found even when the interaction in some subsystems is repulsive. Sometimes it is possible to generate a three-body state even when the interaction in all the subsystems is not strong enough to form individual two-body bound states or resonances. Such states are called as borromean states~\cite{borro}. Thus, the interaction between one or two subsystems can be repulsive or weak, however if the dynamics involved in the remanent subsystem(s) is strong enough to overcome the repulsion/weak attraction, a state of a three-body nature can be formed. This is, indeed, the case of the $KK\bar K$, $\phi K\bar K$, $J/\psi K\bar K$ systems and three-body bound states or resonances are found and associated with the $K(1460)$, $\phi(2170)$ and $Y(4260)$ states, respectively~\cite{mko1,mko2,mko3}. 

The possibility of finding a three-body state in the $\eta^\prime K\bar K$ system has actually been studied earlier in Refs.~\cite{albaladejo, liang}, but conclusions opposite to each other have been found. While in Ref.~\cite{albaladejo}, when the $\eta^\prime K\bar K$ system rearranges as a $\eta^\prime$ and the $f_0(980)$ resonance, a state is found at 1835 MeV with a width of 70 MeV, no signal of such a state is found in Ref.~\cite{liang}. The main difference between the two works is the way of dealing with the three-body dynamics. In Ref.~\cite{albaladejo}, for studying the interaction between $\eta^\prime$ and $f_0(980)$, loops involving these two mesons are introduced and regularized using the dimensional regularization scheme. This implies the introduction of a subtraction constant in the loop function related to the propagation of a meson ($\eta^\prime$) and a resonance [$f_0(980)$]. In Ref.~\cite{liang}, the formation of states in the $\eta^\prime K\bar K$ system is studied within the Faddeev equations in the fixed center approximation approach. In this case, it is assumed that when the $\eta^\prime$ meson interacts with the $K\bar K$ system, which is considered to cluster as the $f_0(980)$ resonance, no changes are produced on the latter.  
The description of the dynamics in the cluster is introduced through a form factor which depends on the mass and width of the cluster~\cite{fca}.

In this paper, we study the $\eta^\prime K\bar K$ system by solving the Faddeev equations with the purpose of looking for possible bound states or/and resonances. We do not assume any cluster formation which cannot be excited in the intermediate scattering states. Such contribution can be important, as noted in Ref.~\cite{javi}.
We obtain the scattering $T$-matrix of the three-body system as a sum of the Faddeev partitions~\cite{faddeev1}, $T_i$, such that
\begin{align}
T=\sum_{i=1}^3 T^i.
\end{align}
 The formalism used here was developed in Refs.~\cite{mko1,mko2,mko3}. As shown in these latter works, the $T_i$ partitions can be rewritten as
\begin{align}
T^i=t^i\delta^3(\vec{k}^\prime_i-\vec{k}_i)+\sum_{j\neq i=1} ^{3}T_R^{ij},
\end{align}
where $T_R^{ij}$ satisfy the equations
\begin{align}
T_R^{ij}=t^i g^{ij}t^j+t^i\left[G^{iji}T^{ji}_R+G^{ijk}T^{jk}_R\right],\label{TR}
\end{align}
for  $i\ne j, j\ne k = 1,2,3$. In Eq.~(\ref{TR}), the function $g^{ij}$ is the three-body Green's function of the system, which is defined as
\begin{eqnarray}
\lefteqn{
g^{ij} (\vec{k}^\prime_i, \vec{k}_j)=\Bigg(\frac{N_{k}}{2E_k(\vec{k}^\prime_i+\vec{k}_j)}\Bigg)}&& \nonumber \\
&& \times \frac{1}{\sqrt{s}-E_i
(\vec{k}^\prime_i)-E_j(\vec{k}_j)-E_k(\vec{k}^\prime_i+\vec{k}_j)+i\epsilon},
\end{eqnarray}
where $\sqrt{s}$ is the energy in the center of mass of the system, the coefficient $N_{k}$ is equal to 1 for mesons and $E_{l}$ ($l=1,\,2,\,3$) is the energy for the particle $l$.

The $G^{ijk}$ function in Eq.~(\ref{TR}) represents  a loop function of three-particles and it is written as
\begin{equation}
G^{i\,j\,k}  =\int\frac{d^3 k^{\prime\prime}}{(2\pi)^3}\tilde{g}^{ij} \cdot F^{i\,j \,k},
\label{eq:Gfunc}
\end{equation}
with the elements of  $\tilde{g}^{ij}$ being 
\begin{eqnarray}
\lefteqn{
\tilde{g}^{ij} (\vec{k}^{\prime \prime}, s_{lm}) = \frac{N_l}
{2E_l(\vec{k}^{\prime\prime})} \frac{N_m}{2E_m(\vec{k}^{\prime\prime})} } && \nonumber \\
&& \quad \times
\frac{1}{\sqrt{s_{lm}}-E_l(\vec{k}^{\prime\prime})-E_m(\vec{k}^{\prime\prime})
+i\epsilon},
\label{eq:G} 
\end{eqnarray}
for $i \ne l \ne m$, 
and the $F^{i\,j\,k}$ function, with explicit variable dependence, is given by 
\begin{eqnarray}
\lefteqn{
F^{i\,j\,k} (\vec{k}^{\prime \prime},\vec{k}^\prime_j, \vec{k}_k,  s^{k^{\prime\prime}}_{ru})=  } && \nonumber \\
&& t^{j}(s^{k^{\prime\prime}}_{ru}) g^{jk}(\vec{k}^{\prime\prime}, \vec{k}_k)
\Big[g^{jk}(\vec{k}^\prime_j, \vec{k}_k) \Big]^{-1}
\Big[ t^{j} (s_{ru}) \Big]^{-1},  \label{offac}
\end{eqnarray}
for $ j\ne r\ne u=1,2,3$.
In Eq. (\ref{eq:G}), $\sqrt{s_{lm}}$ is the invariant mass of the $(lm)$ pair and it depends on the external variables. The upper index $k^{\prime\prime}$ in the invariant mass $s^{k^{\prime\prime}}_{ru}$ of Eq.~(\ref{offac}) indicates its dependence on the loop variable 
(see Refs.~\cite{mko1,mko2,mko3,mko4,mj} for more details). 

The input two-body $t$-matrices of Eq.~(\ref{TR}) are obtained by solving the Bethe-Salpeter equation in a coupled channel approach
\begin{align}
t&=V+VGt,\label{bet}\\ 
&=V + \int\frac{d^4 k}{(2\pi)^4}V\frac{1}{[(P-k)^2-m^2_1+i\epsilon][k^2-m^2_2+i\epsilon]}t,\nonumber
\end{align}
where the kernel $V$ is determined from the Lagrangian given by Eq.~(\ref{L}). The  $G$ function in Eq.~(\ref{bet}) stands for the two-body loop function, $P$ and $k$ are, respectively, the total four momentum of the two body system and that of the particles in the loop (expressed in the two-body center of mass frame), and $m_1$ and $m_2$ the masses of the two particles under consideration. 

The first step of our formalism is to solve Eq.~(\ref{bet}) for all the two-body subsystems by considering all the relevant coupled channels into account. In this way, the resonances generated in the two-body subsystems are automatically present in the three-body scattering. 

As shown in Refs.~\cite{oset,pelaez,oller4,hyodo}, it is possible to convert the integral Bethe-Salpeter equation [Eq.~(\ref{bet})] into algebraic equations. In this case, the kernel $V$, and thus, $t$, can be factorized outside the integral and Eq.~(\ref{bet}) becomes
 \begin{align}
 t=[1-VG]^{-1}V,\label{B2}
 \end{align}
where the loop function $G$ is regularized using dimensional regularization or a cut-off~\cite{oset,pelaez}. 
 
In a similar fashion, as shown in Refs.~\cite{mko1,mj}, equation~(\ref{TR}) is also an algebraic set of six coupled equations. This simplification is a result of the cancellation of the contribution of the off-shell parts of the two-body $t$-matrices in the three-body Faddeev partitions with the  contact term(s) of same topology (whose origin relies in the Lagrangian used to describe the two-body interaction in the subsystems)~\cite{mko1,mko2,mko3,mko4,mj}. Interestingly,  a deduction of cancellations of two-body and three-body forces using a different procedure has recently been reported in Ref.~\cite{arriola}. Due to these cancellations, only the on-shell part of the two-body $t$-matrices is relevant to solve Eq.~(\ref{TR}). As a consequence, the $T^{ij}_R$ partitions given in Eq.~(\ref{TR}) depend only on the total three-body energy, $\sqrt s$, and on the invariant mass of one of the subsystems, which we choose to be the one related to particles 2 and 3 and denote the invariant mass as $\sqrt{s_{23}}$. The other invariant masses, $\sqrt{s_{12}}$ and $\sqrt{s_{31}}$ can be obtained in terms of  $\sqrt s$ and $\sqrt{s_{23}}$, as shown in Refs. \cite{mko2,mko3}.

Using this formalism, we solve Eq.~(\ref{TR}) for the $\eta^\prime K\bar K$ system. The input two-body $\eta^\prime K$ and $\eta^\prime\bar K$ amplitudes are obtained following Ref.~\cite{guo}, where Eq.~(\ref{bet}) is solved for the $\pi K$, $\eta K$ and $\eta^\prime K$ system and, as a result of this coupled channel dynamics, the $\kappa$ resonance is generated. A good reproduction of the $\pi K$ phase shift is found up to energies slightly above 1.3 GeV.~For the $K\bar K$ $t$-matrix we consider the work of Ref.~\cite{oset}, in which the $\pi\pi$, $K\bar K$ system is investigated for the isospin 0 configuration and, for the isospin 1 case, the $K\bar K$ and $\pi\eta$ system is considered. Due to the dynamics involved in these coupled channel systems, $f_0 (600)$ and $f_0(980)$ are found for the isospin 0 case and $a_0(980)$ for the isospin 1 case. The experimental $\pi\pi$ phase-shifts are well reproduced up to energies around 1.2 GeV.  

\begin{figure*}[t]
\centering
\includegraphics[width=\textwidth]{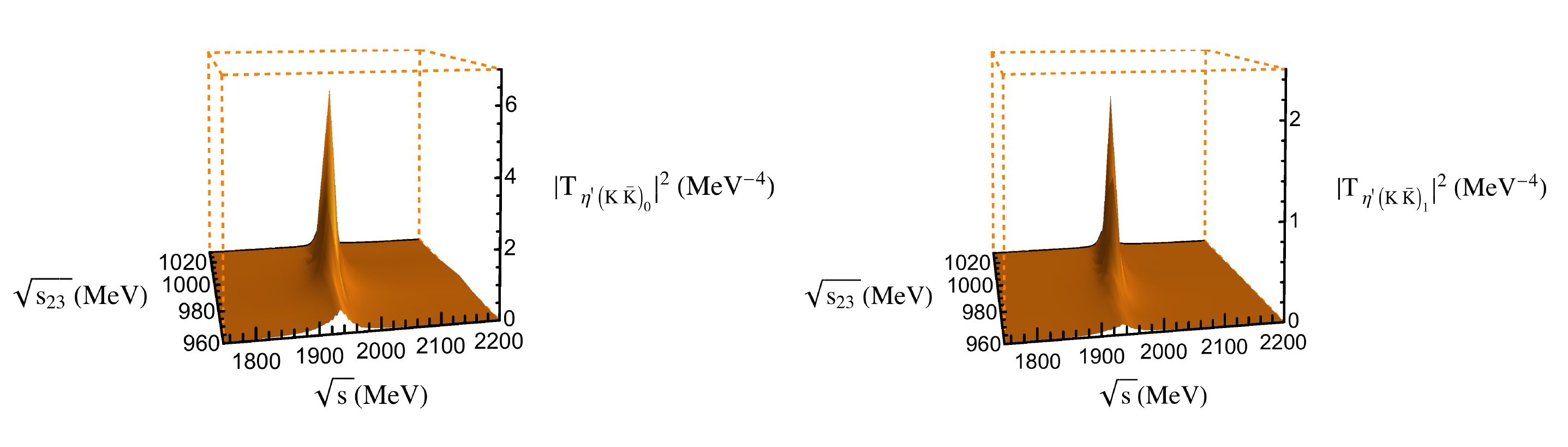}
\caption{Modulus squared of the three-body $T$-matrix for the $\eta^\prime K\bar K$ system for total isospin 0, thus the $K\bar K$ subsystem is in isospin 0 (left panel) and for total isospin 1, which implies that the $K\bar K$ subsystem is in isospin 1 (right panel). The peak seen in both figures corresponds to the three-body threshold cusp.
 }\label{hf0}
\end{figure*}

In Fig.~\ref{hf0} 
%\begin{widetext}
%\end{widetext}
we show the plots obtained for the $\eta^\prime K\bar K$ $T$-matrix for total isospin 0 (left panel) and 1 (right panel) as a function of $\sqrt{s}$ and $\sqrt{s}_{23}$. As can be seen, apart from the threshold enhancement at $(\sqrt{s},\sqrt{s}_{23})=(1960, 992)$ MeV in both isospins, no other structure is found. Not even for values of $\sqrt{s}_{23}$ around 980 MeV, where the $K\bar K$ system in isospin 0 forms $f_0(980)$ and in isospin 1 forms $a_0(980)$. A threshold enhancement was also the only effect seen in the study of Ref.~\cite{liang}. At this point a question might arise about the stability of our results when the subtraction constants/cut-offs of the loop functions are varied. In the case of the calculation of the two-body $t$-matrices, the subtraction constants/cut-offs used here, following Refs.~\cite{guo} and \cite{oset}, have been fixed to reproduce relevant data on phase-shifts and inelasticities. We have not varied them due to the limited availability of freedom. For the three-body loop functions, Eq.~(\ref{eq:Gfunc}), a cut-off of 1000 MeV has been used. We have varied this cut-off in the range 800-1100 MeV, and minor changes in the size of the three-body amplitudes of Fig.~\ref{hf0} are observed. This insensitivity is related to the presence of three-meson propagators in Eq.~(\ref{eq:Gfunc}).
Thus, our study of the $\eta^\prime K\bar K$ system reveals no structure neither at 1835 MeV, contrary to the finding of Ref.~\cite{albaladejo}, nor above the threshold. Hence, we cannot relate $X(1835)$ and  $X(2120)$ with states generated by three-body dynamics. The third $X$ found in Ref.~\cite{Ablikim:2010au},  $X(2370)$, is anyways too heavy to be explained as as $\eta^\prime K \bar K$ resonance.  Apart from $X(1835)$, $X(2120)$, there are some $\pi, \eta$ states listed in the PDG at energies 1800-2100 MeV with large widths, 100-200 MeV: $\eta(1760)$, $\pi(1800)$, $\eta(2225)$. According to the study carried in this work, the dynamics involved in the $\eta^\prime K\bar K$ system plays no essential role in understanding the nature of the above mentioned states. We thus conclude from our work that the origin of $X(1835)$ and  $X(2120)$ must be something other than three-pseudoscalar dynamics.

We thank Professors Jos\'e Antonio Oller and Eulogio Oset for reading the manuscript and for useful discussions.
%\clearpage

\end{document}